\documentclass{ws-procs9x6}

\begin{document}

\title{MEASUREMENT OF THE SCINTILLATION EFFICIENCY OF Na RECOILS IN NaI(Tl)
  DOWN TO 10 keV NUCLEAR RECOIL ENERGY RELEVANT TO DARK MATTER SEARCHES}

\author{H.~CHAGANI$^*$, P.~MAJEWSKI$^{**}$, E.~J.~DAW, V.~A.~KUDRYAVTSEV, and
  N.~J.~C.~SPOONER}

\address{Department of Physics and Astronomy, University of Sheffield, Hicks
  Building, Hounsfield Road, Sheffield S3 7RH, United Kingdom \\
  $^*$E-mail: h.chagani@sheffield.ac.uk \\
  $^{**}$E-mail: p.majewski@sheffield.ac.uk}

\begin{abstract}
We present preliminary results of measurements of the quenching factor for Na
recoils in NaI(Tl) at room temperature, made at a dedicated neutron facility at
the University of Sheffield. Measurements have been performed with a 2.45~MeV
mono-energetic neutron generator in the energy range from 10~keV to 100~keV
nuclear recoil energy. A BC501A liquid scintillator detector was used to tag
neutrons. Cuts on pulse-shape discrimination from the BC501A liquid
scintillator detector and neutron time-of-flight were performed on pulses
recorded by a digitizer with a 2~ns sampling time. Measured quenching factors
range from 19\%~to 26\%, in agreement with other experiments. From pulse-shape
analysis, a mean time of pulses from electron and nuclear recoils are compared
down to 2~keV electron equivalent energy.
\end{abstract}

\bodymatter

\section{Introduction}
Thallium activated sodium iodide (NaI(Tl)) crystals are a popular choice as a
target material for dark matter experiments. This is because of their
relatively high light yield and pulse shape differences between nuclear and
electron recoils. Currently three experiments utilise these crystals:
ANAIS~\cite{Martinez06}; DAMA/NaI~\cite{Bernabei98}; and
ELEGANT-V~\cite{Fushimi99}. Although better discrimination can be reached with
other target media, NaI(Tl) crystal-based detectors remain one of the best at
determining spin-dependent WIMP-nucleon limits. For instance, for direct
detection techniques, the NaIAD experiment~\cite{Ahmed03} still holds the best
spin-dependent limit on WIMP-proton interactions~\cite{Alner05}. Hence, NaI(Tl)
remains an important detector material in non-baryonic dark matter searches.

\section{Measuring Scintillation Efficiencies}
An important measurement to determine a scintillating target's sensitivity to
dark matter particles is the ratio of light induced by a nuclear recoil
(produced by WIMP/neutron collisions with nuclei) to that by an electron recoil
of the same energy. This is known as the quenching factor.

\begin{figure}
  \begin{center}
    \includegraphics[scale=0.59,angle=270]{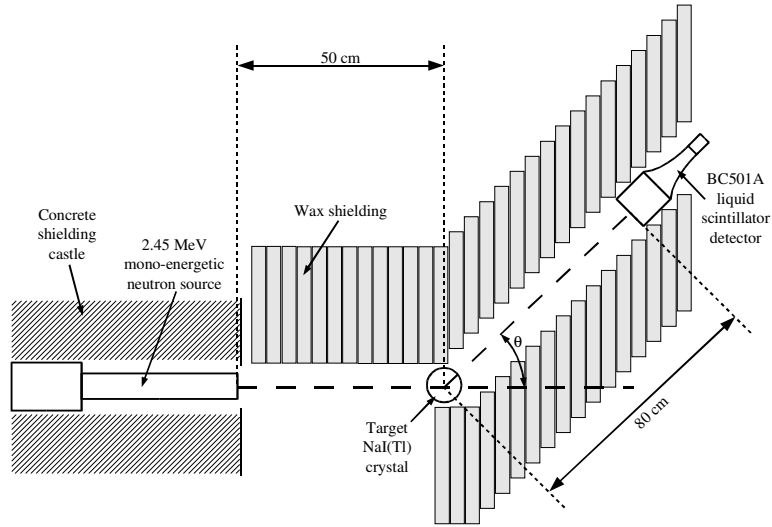}
    \caption{Schematic view of the detector arrangement used to measure
      the scintillation efficiency of nuclear recoils in NaI(Tl).}
    \label{qf_setup}
  \end{center}
\end{figure}

The 2-inch diameter, encapsulated, NaI(Tl) crystal is placed in the path of a
2.45~MeV mono-energetic deuterium-deuterium neutron beam as shown in
Figure~\ref{qf_setup}. A 3-inch ETL 9265KB photomultiplier tube is optically
coupled to the crystal. Neutrons that interact with the target nuclei scatter
off at an angle that is dependent on the deposited neutron energy:

\begin{equation}\label{scatter_eq}
E_{R} \approx \frac{2m_{A}E_{n}m_{n}}{(m_{A} + m_{n})^{2}} \cdot
(1 - \cos \theta)
\end{equation}
where $E_{R}$ is the recoil energy, $m_{A}$ is the mass of the target nucleus,
$E_{n}$ is the energy of the neutron beam, $\theta$ is the scattering angle and
$m_{n}$ is the mass of the neutron.

A BC501A liquid scintillator detector is placed at various angles around the
target to detect scattered neutrons at nuclear recoil energies given by
Eq.~(\ref{scatter_eq}). Pulses from the target and liquid scintillator that are
coincident within a 100~ns time window are sent to a 2-channel, 8-bit, Acqiris
digitizer with a 500~MHz sampling rate, and written to disk.

\section{NaI(Tl) Response to Electron Recoils}
\begin{figure}
  \begin{center}
    $\begin{array}{c@{\hspace{5mm}}c}
      \includegraphics[scale=0.26]{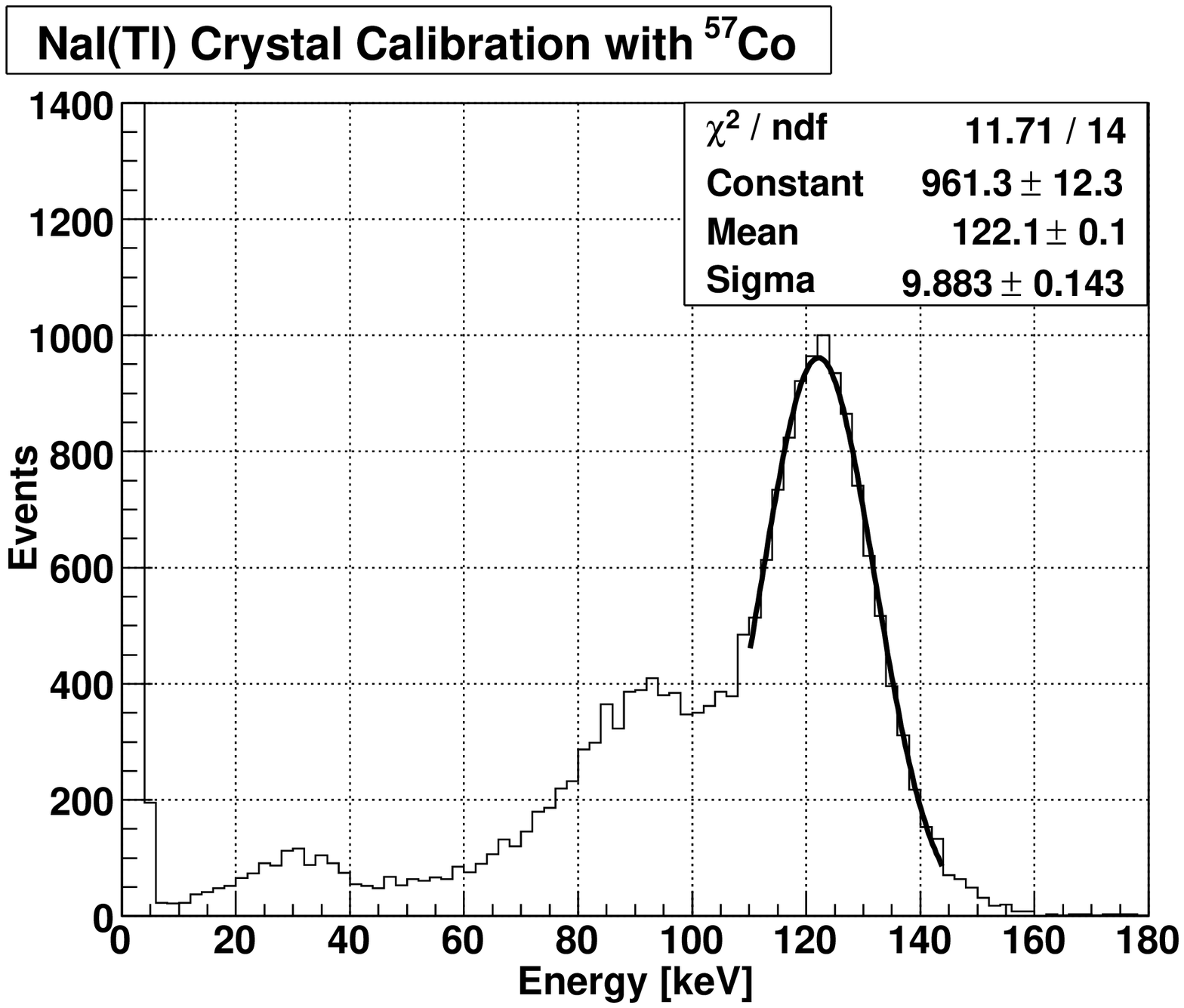} &
      \includegraphics[scale=0.26]{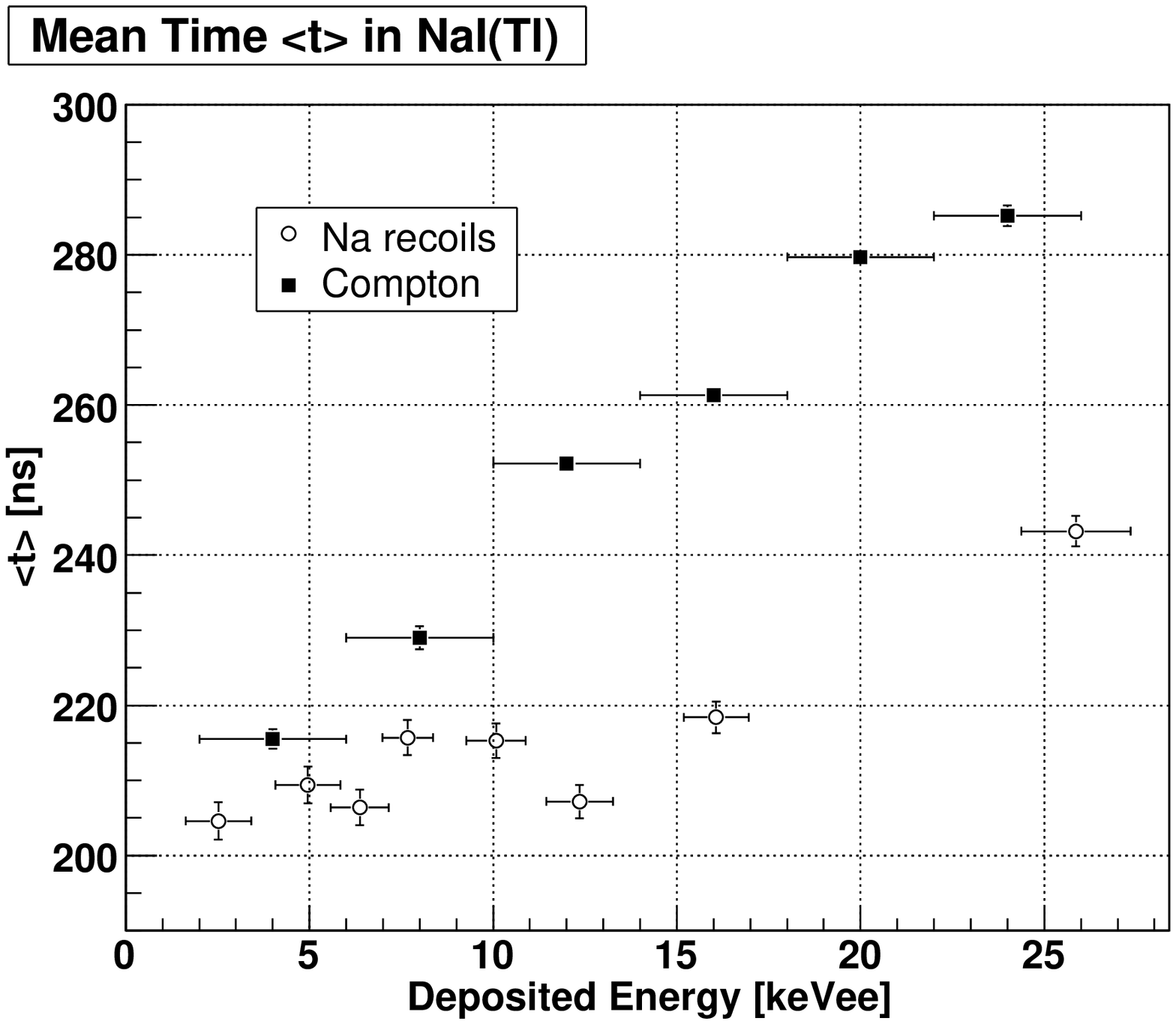} \\
      \mbox{\small{(a)}} & \mbox{\small{(b)}}
    \end{array}$
    \caption{(a)~Results from calibration of detector with 122~keV line from
      $^{57}\mbox{Co}$ source. (b)~Mean time of the pulses as a function of
      deposited energy for sodium (Na) recoils and Compton electrons.
      Measurements of Na recoils are performed with the neutron beam. Compton
      electrons are induced by a $^{22}\mbox{Na}$ source.}
    \label{response}
  \end{center}
\end{figure}

It is known that the response of NaI(Tl) to gamma-rays is non-linear at low
energies~\cite{Gerbier99,Wayne98}. The crystal is exposed to gamma-rays from a
variety of low energy sources, between 29~keV ($^{129}\mbox{I}$ Xe
$\mbox{K}_{\alpha}$ X-rays) and 122~keV ($^{57}\mbox{Co}$ $\gamma$-rays). A
decrease in detector response is observed at the iodine K-shell absorption edge
at 33.2~keV, consistent with other studies~\cite{Gerbier99,Wayne98}. Therefore,
determination of the energy scale must be performed in an energy region where a
linear response is observed. The crystal is calibrated with a $^{57}\mbox{Co}$
source, as shown in Figure~\ref{response}(a). The iodine escape peak at 90~keV
is also seen. A light yield of 5.5~photoelectrons/keV is found.

\vspace{-7mm}
\section{Quenching Factor Measurements}
The nuclear recoil energy spectrum measured at each scattering angle is
converted to an electron equivalent energy scale (keVee) and compared to the
expected nuclear recoil energy (keVnr) at that angle. The ratio between the
measured nuclear recoil energy, and that calculated from
Eq.~(\ref{scatter_eq}), is the quenching factor.

In order to eliminate the background from gamma-ray interactions, nuclear
recoils are discerned by discrimination of pulse shapes from the BC501A
detector and time-of-flight measurements.

\begin{figure}
  \begin{center}
    $\begin{array}{c@{\hspace{5mm}}c}
	\includegraphics[scale=0.27]{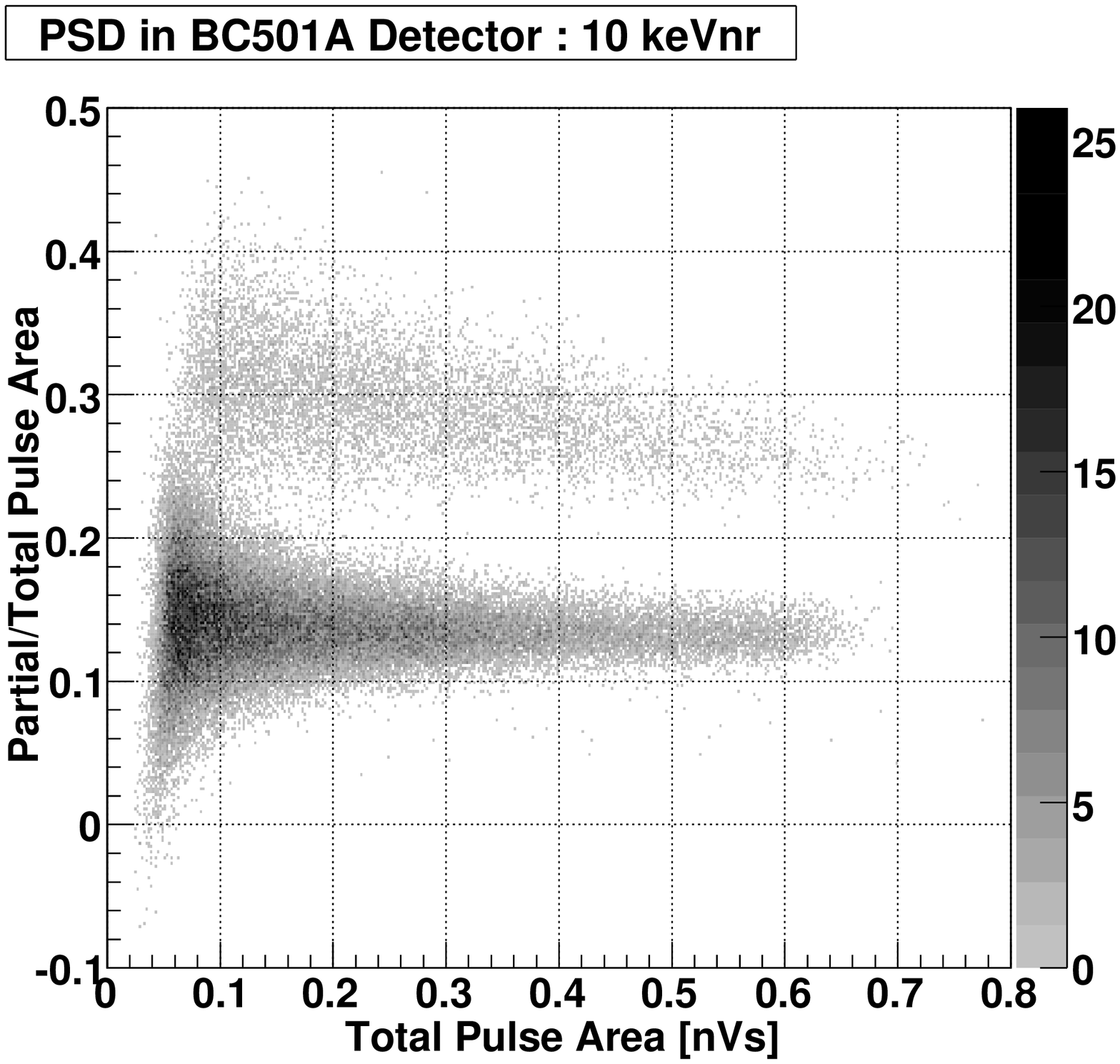} &
	\includegraphics[scale=0.27]{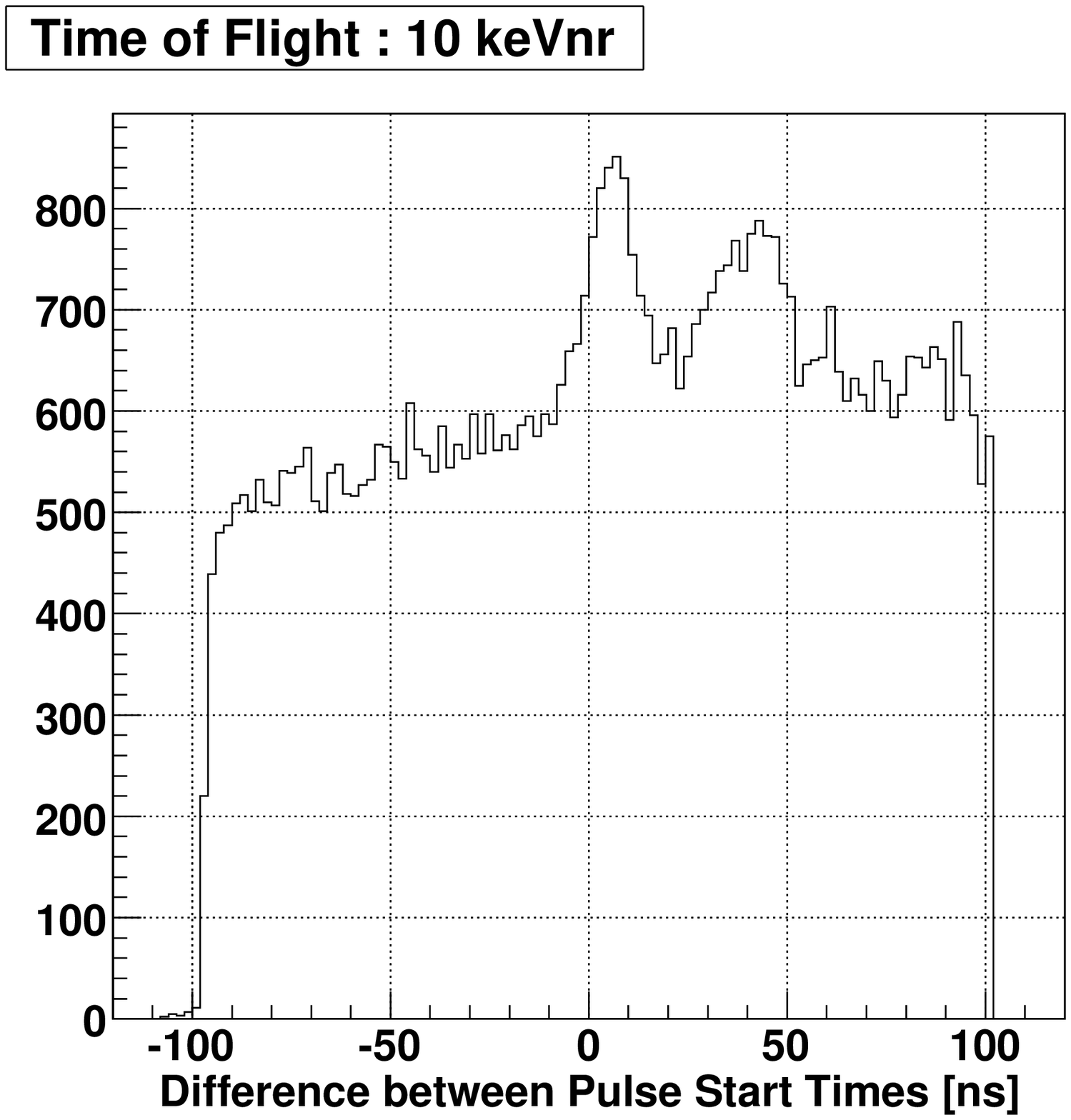} \\
	\mbox{\small{(a)}} & \mbox{\small{(b)}} \\ \\
	\includegraphics[scale=0.27]{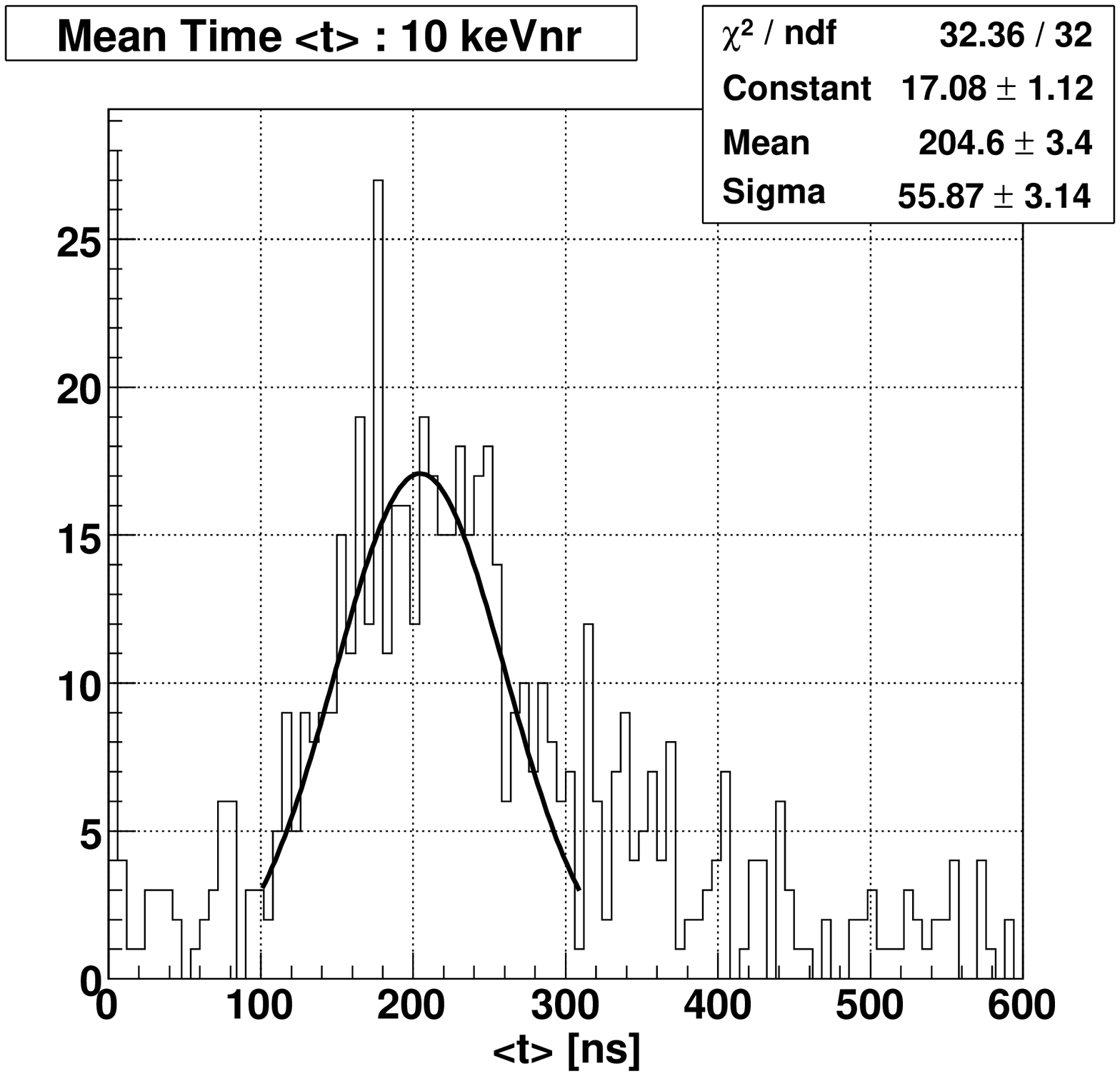} &
	\includegraphics[scale=0.27]{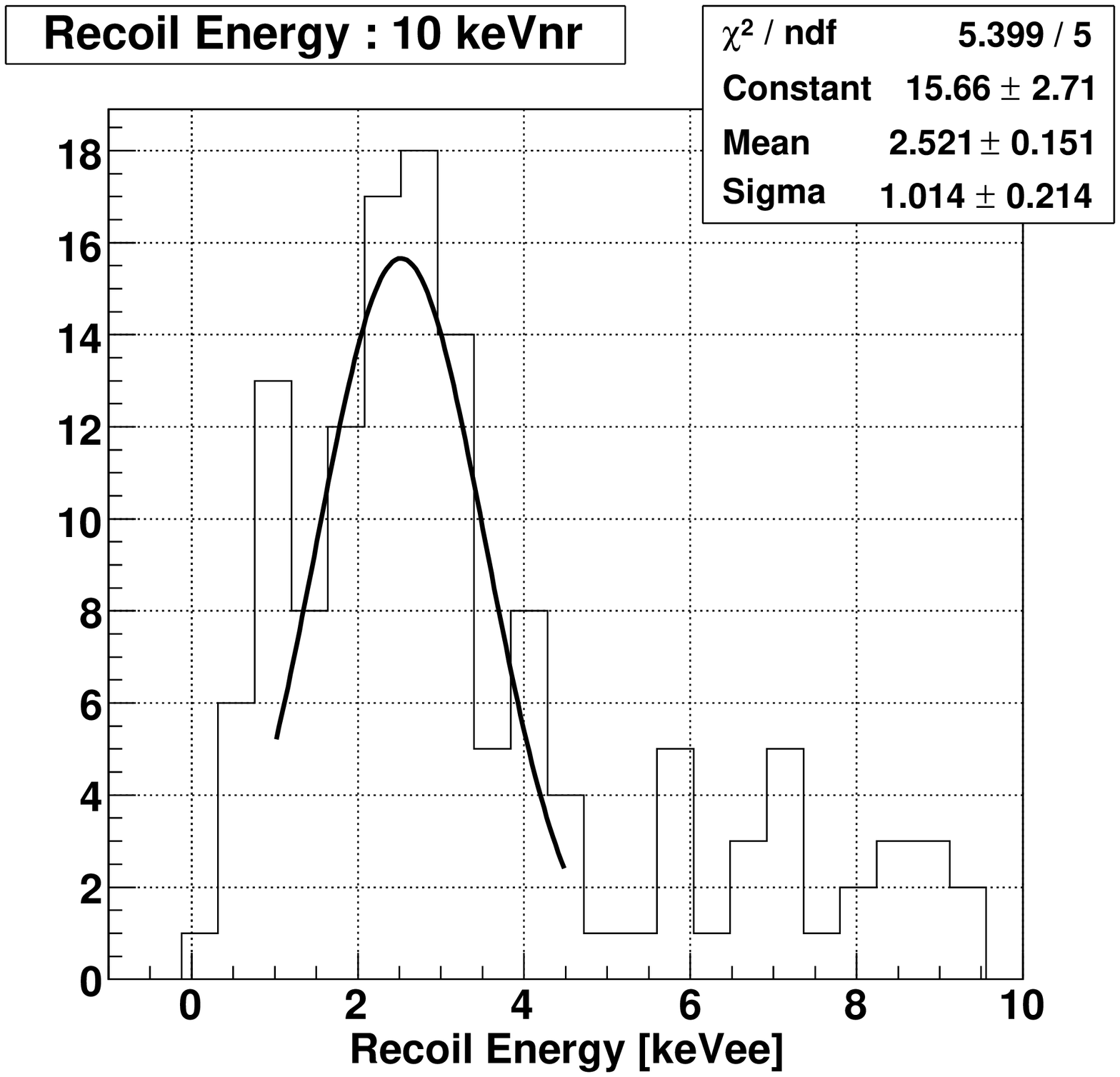} \\
	\mbox{\small{(c)}} & \mbox{\small{(d)}}
    \end{array}$
    \caption{(a)~Pulse shape discrimination in the BC501A liquid scintillator
      detector for 10~keVnr Na recoils. The upper neutron event band is clearly
      distinguishable from gamma interactions. (b)~Time-of-flight between the
      BC501A detector and NaI(Tl) crystal. The peaks at roughly 0~and 40~ns
      represent gamma and neutron events respectively. (c)~Mean time of pulses
      from 10 keVnr Na recoils in the NaI(Tl) crystal. (d)~Recoil energy in
      electron equivalent scale.}
    \label{psd}
  \end{center}
\end{figure}

Events that arise from gamma-ray interactions in the BC501A detector have
shorter decay times relative to those from neutron collisions. Adequate
discrimination between nuclear and electron recoils is achieved by integrating
over the tail of the pulse. The ratio between this partial area and the total
pulse area is then plotted as a function of the total area. An example is
illustrated in figure~\ref{psd}(a).

Unlike gammas, neutrons that scatter off the target in this experiment are
non-relativistic. Therefore, after interacting with a sodium nucleus in the
crystal, a neutron takes approximately 40~ns to travel the 80~cm distance to
the BC501A detector. By taking the time difference between coincident events in
the target and BC501A detector, as shown in Figure~\ref{psd}(b), neutron and
gamma events are separated.

Better discrimination at energy scales relevant to dark matter searches has
been demonstrated in NaI(Tl) crystals using the mean time
$\langle t\rangle$~\cite{Gerbier99,Kudryavtsev99} rather than the traditional
double charge method~\cite{Morris76}. The mean time is defined as:
$\langle t\rangle = \frac{\sum_{i} A_{i} t_{i}}{\sum_{i} A_{i}}$, where $A_{i}$
is the amplitude of the digitized pulse at the time bin $t_{i}$. After gamma
events have been rejected by performing the cuts outlined above, mean time
distributions for each nuclear recoil energy are investigated. An example is
shown in Figure~\ref{psd}(c). This is compared with the mean time of Compton
recoils induced by a $^{22}\mbox{Na}$ source in Figure~\ref{response}(b). In
agreement with previous work~\cite{Tovey98}, it is clear that this difference
is less prominent at energies less than 10~keVee, limiting the discrimination
power of NaI(Tl) detectors. This cut serves to reduce the background from noise
and gamma pulses, leading to a nuclear recoil peak as shown in
Figure~\ref{psd}(d).

\section{Results}
\begin{figure}
  \begin{center}
    \includegraphics[scale=0.49]{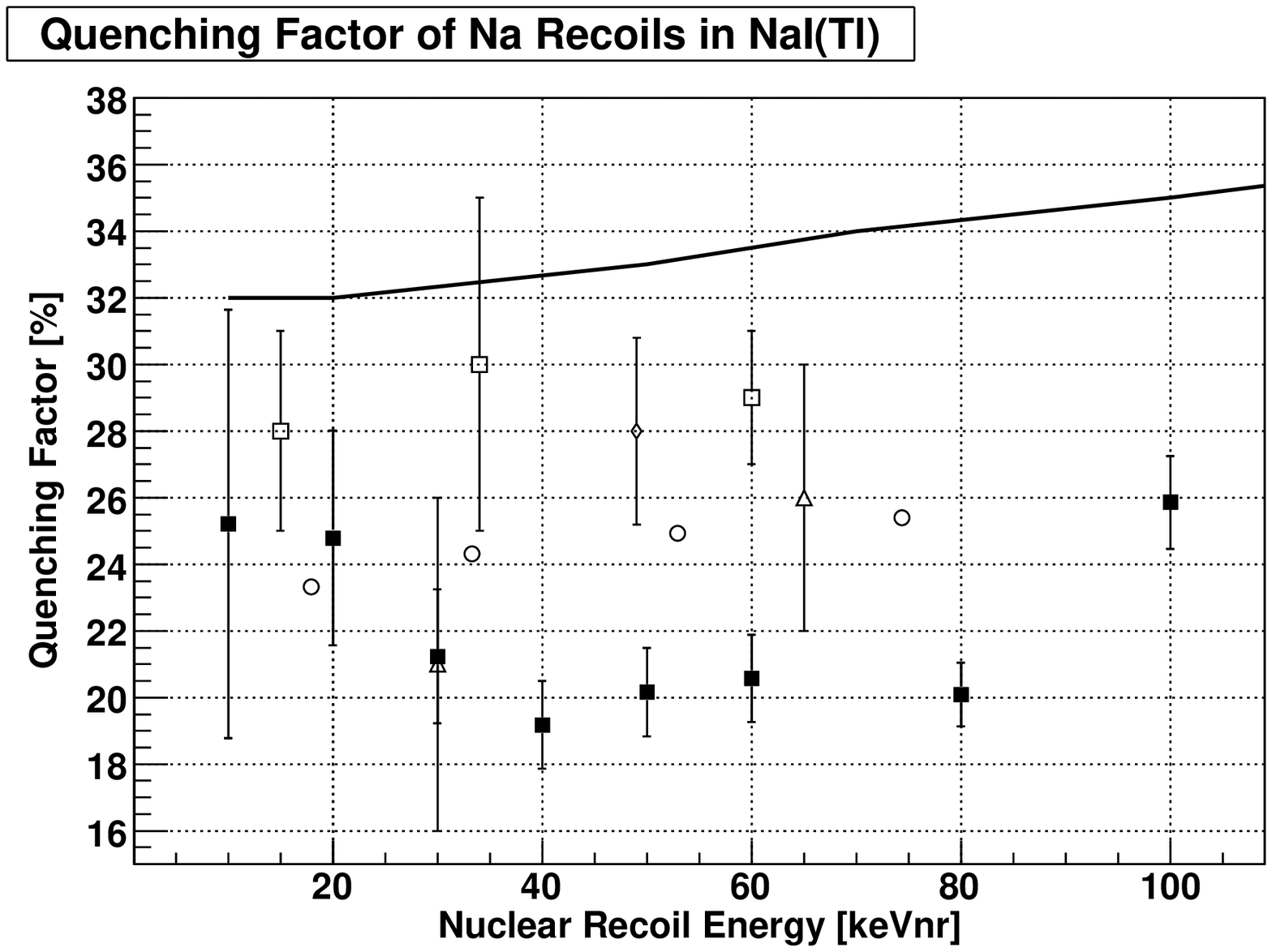}
    \caption{Quenching factor measurements for Na recoils in NaI(Tl). Results
      from this work (closed black squares), Simon et al.~\cite{Simon03} (open
      diamond), Gerbier et al.~\cite{Gerbier99} (open circles), Tovey et
      al.~\cite{Tovey98} (open triangles), Spooner et al.~\cite{Spooner94}
      (open squares) and the preliminary theoretical estimation from
      Hitachi~\cite{Hitachi06} (solid black line) are shown.}
    \label{QF_plot}
  \end{center}
\end{figure}

The quenching factor varies between 19\% to 26\% in the range 10~keV to 100~keV
nuclear recoil energy, which agrees with previous experimental
results~\cite{Gerbier99,Tovey98,Simon03,Spooner94}, as shown in
Figure~\ref{QF_plot}. From Figure~\ref{psd}(d), a scintillation efficiency of
$25.2 \pm 6.4$\% has been determined for 10~keVnr Na recoils.

\section{Summary}
Scintillation efficiency measurements have been performed for Na recoils in a
NaI(Tl) crystal. The results show an average value of 22.1\% at energies less
than 50~keVnr, and are in agreement with other measurements. Future plans
include simulating nuclear recoils in NaI(Tl) in an effort to decrease the
errors in quenching factors.

\section{Acknowledgements}
The authors would like to thank Akira Hitachi for valuable discussions of the
results, and the provision of a preliminary theoretical estimation of the
quenching factor. HC would also like to thank PPARC for a PhD studentship.

\end{document}